\documentclass[aps, prd, amsmath, floats, floatfix, twocolumn, nofootinbib,
 superscriptaddress, showpacs]{revtex4-1}
\usepackage{graphicx}
\usepackage{color}
\usepackage[normalem]{ulem}

\newcommand{\be}{\begin{eqnarray}}
\newcommand{\ee}{\end{eqnarray}}

\begin{document}

\title{Testing the existence of regions of stable orbits at small radii around black hole candidates}

\author{Cosimo Bambi}
\email{bambi@fudan.edu.cn}
\affiliation{Center for Field Theory and Particle Physics \& Department of Physics,
 Fudan University, 200433 Shanghai, China}

\author{Georgios Lukes-Gerakopoulos}
\email{gglukes@gmail.com}
\affiliation{Theoretical Physics Institute, Friedrich-Schiller-Universit\"at Jena,
 D-07743 Jena, Germany}

\date{\today}

\begin{abstract}
Black hole candidates in X-ray binary systems and at the centers of galaxies are 
expected to be the Kerr black holes of General Relativity, but the actual nature 
of these objects has yet to be verified. In this paper, we consider the possibility that
they are exotic compact objects and we describe their exterior gravitational field 
with a subclass of the Manko-Novikov metrics, which are exact solutions of the 
vacuum Einstein's equations and can describe the spacetime geometry around bodies with 
arbitrary mass-multipole moments. We point out that around a Manko-Novikov object 
there may exist many disconnected non-plunging regions at small radii, with no 
counterpart in the Kerr background, and that their existence may be tested. 
For instance, in the presence of an accretion disk, they may be filled by the accreting 
gas, forming a ring structure that might remind the rings of Saturn. We suggest that 
the existence of these regions may have a clear observational signature in the 
waveform of the gravitational radiation emitted by an EMRI: in the last stage of 
the inspiral, the waveform would be the combination of ``regular chirps'', produced 
when the small object orbits in one of the non-plunging regions, and ``bursts'', 
released when the small object jumps from a non-plunging region to another 
one at smaller radii. Our conclusions are supported by some 
numerical calculations of trajectories in the geodesic approximation, in which 
a particle plunges from the ISCO and then seems to get trapped in the potential 
well at smaller radii.
\end{abstract}

\pacs{97.60.Lf, 04.50.Kd, 97.10.Gz}

\maketitle


\section{Introduction}

In 4-dimensional General Relativity, uncharged black holes (BHs) are described by
the Kerr solution and are characterized by only two parameters: the mass, $M$, and
the spin angular momentum, $J$~\cite{hair}. Astrophysical BH candidates are
stellar-mass compact objects in X-ray binary systems~\cite{bh1} and super-massive
bodies at the centers of galaxies~\cite{bh2}\footnote{The existence of a third class
of objects, intermediate-mass  BH candidates with $M \sim 10^2 - 10^4$~$M_\odot$,
is still controversial, because their detections are indirect, and definitive
dynamical measurements of their masses  are lacking~\cite{bh5}.}. The former have a
mass exceeding 3~$M_\odot$ and are therefore too heavy to be neutron or quark stars
for any plausible matter equation of state~\cite{bh3}. At least some of the
super-massive BH candidates in galactic nuclei turn out to be too heavy, compact,
and old to be clusters of non-luminous bodies~\cite{bh4}. These two classes of
objects are thought to be the Kerr BH predicted by General Relativity because there
is no alternative explanation in the framework of conventional physics. However,
their actual nature has yet to be proven~\cite{review}.

In this work, we consider the possibility that astrophysical BH candidates are exotic 
compact objects and we describe their {\it exterior} gravitational field with a subclass 
of the Manko-Novikov (MN) metrics~\cite{mn}. The MN spacetimes are exact solutions
of the vacuum  Einstein's equations and they are characterized by the mass of the 
compact object, its  spin angular momentum, and an infinite number of deformation
parameters, setting all the mass-multipole moments. Previous work on these metrics
considered only the simplest case with one deformation parameter, the anomalous
mass-quadrupole moment $q$, assuming zero all the higher anomalous mass 
moments~\cite{mn-glm,mn-glg,mn-cb,mn-cbeb}. The existence of some disconnected 
non-plunging regions at small radii in the MN spacetimes, which have no counterpart 
in the Kerr BH background (but see~\cite{pqr}), is already known. So far they 
have been quite ignored and were considered to be of no 
astrophysical interest. The reason is that it was thought that
their orbital energy is always higher than the one at larger radii and therefore 
that they would be difficult to populate by the gas of the accretion disk or by small 
bodies inspiralling into the BH candidate. Here we point out that this is not always
true and that the existence of such non-plunging regions may have astrophysical 
implications. Moreover, in addition to a non-vanishing $q$, we consider the possibility 
of a non-vanishing anomalous mass-hexadecapole moment $h$, and we argue 
that the maximum number of non-plunging regions increases as the number 
of non-zero deformation parameters increases.

In the Kerr metric, equatorial circular orbits are always vertically stable, while
they are radially stable for radii $r > r_{\rm ISCO}$, where $r_{\rm ISCO}$ is the
radius of the innermost stable circular orbit (ISCO). Equatorial circular orbits with 
radii $r_{\rm PO}< r < r_{\rm ISCO}$, where $r_{\rm PO}$ is the radius of the photon
orbit, are unstable for small perturbations along the radial direction, while there
are no equatorial circular orbits for $r < r_{\rm PO}$. In non-Kerr spacetimes,
equatorial circular orbits may also be vertically unstable, which leads to a number
of completely new phenomena~\cite{mn-cbeb,chl}. In the MN metric with a
non-vanishing anomalous quadrupole moment, there may be an island of stable
equatorial circular orbits in the region $r < r_{\rm ISCO}$; that is, there may
exist two radii $r_1$ and $r_2$ such that equatorial circular orbits with
$r_1 < r < r_2$ are stable. If we consider the MN metric with non-vanishing
anomalous mass-quadrupole and mass-hexadecapole moment, there may exist 
two disconnected non-plunging regions for $r < r_{\rm ISCO}$, say $r_1 < r < r_2$ and 
$r_4 < r < r_5$. In the presence of an accretion disk, these regions may be filled 
by the accreting gas, because their energy and angular momentum can be 
lower than the ones at larger radii.

Throughout the paper, we use units in which $G_{\rm N} = c = 1$.

\section{Manko-Novikov spacetimes}

The MN metrics~\cite{mn} are stationary, axisymmetric, and asymptotically flat
exact solutions of the vacuum Einstein's equations with arbitrary mass-multipole 
moments. In quasi-cylindrical coordinates $(\rho,z)$ and in prolate spheroidal 
coordinates $(x,y)$, the line element is, respectively,
\begin{widetext}
\be\label{eq-ds2}
ds^2 &=& - f \left(dt - \omega d\phi\right)^2
+ \frac{e^{2\gamma}}{f} \left(d\rho^2 + dz^2\right)
+ \frac{\rho^2}{f} d\phi^2 = \nonumber\\
&=& - f \left(dt - \omega d\phi\right)^2
+ \frac{k^2 e^{2\gamma}}{f}\left(x^2 - y^2\right)
\left(\frac{dx^2}{x^2 - 1} + \frac{dy^2}{1 - y^2}\right)
+ \frac{k^2}{f} \left(x^2 - 1\right)\left(1 - y^2\right) 
d\phi^2 \, ,
\ee
where
\be
f = e^{2\psi} A/B\, , \quad
\omega = 2 k e^{- 2\psi} C A^{-1} 
- 4 k \alpha \left(1 - \alpha^2\right)^{-1} \, , \quad
e^{2\gamma} &=& e^{2\gamma'}A \left(x^2 - 1\right)^{-1}
\left(1 - \alpha^2\right)^{-2} \, ,
\ee
and
\be
\psi &=& \sum_{n = 1}^{+\infty} \frac{\alpha_n P_n}{R^{n+1}} 
\, , \\\label{gammapdef}
\gamma' &=& \frac{1}{2} \ln\frac{x^2 - 1}{x^2 - y^2} 
+ \sum_{m,n = 1}^{+\infty} \frac{(m+1)(n+1) 
\alpha_m \alpha_n}{(m+n+2) R^{m+n+2}}
\left(P_{m+1} P_{n+1} - P_m P_n\right) + \nonumber\\
&& + \left[ \sum_{n=1}^{+\infty} \alpha_n 
\left((-1)^{n+1} - 1 + \sum_{k = 0}^{n}
\frac{x-y+(-1)^{n-k}(x+y)}{R^{k+1}}P_k \right) \right] \, , \\
A &=& (x^2 - 1)(1 + \tilde{a}\tilde{b})^2 - (1 - y^2)(\tilde{b} - \tilde{a})^2 \, , \\
B &=& [x + 1 + (x - 1)\tilde{a}\tilde{b}]^2 + [(1 + y)\tilde{a} + (1 - y)\tilde{b}]^2 \, , \\
C &=& (x^2 - 1)(1 + \tilde{a}\tilde{b})[\tilde{b} - \tilde{a} - y(\tilde{a} + \tilde{b})] 
+ (1 - y^2)(\tilde{b} - \tilde{a})[1 + \tilde{a}\tilde{b} + x(1 - \tilde{a}\tilde{b})] \, , \\\label{adef}
\tilde{a} &=& -\alpha \exp \left[\sum_{n=1}^{+\infty} 2\alpha_n 
\left(1 - \sum_{k = 0}^{n} \frac{(x - y)}{R^{k+1}} 
P_k\right)\right] \, , \\\label{bdef}
\tilde{b} &=& \alpha \exp \left[\sum_{n=1}^{+\infty} 2\alpha_n 
\left((-1)^n + \sum_{k = 0}^{n} \frac{(-1)^{n-k+1}(x + y)}{R^{k+1}} 
P_k\right)\right] \, .
\ee
\end{widetext}
Here $R = \sqrt{x^2 + y^2 - 1}$ and $P_n$ are the Legendre 
polynomials with argument $xy/R$,
\be
P_n &=& P_n\left(\frac{xy}{R}\right) \, , \nonumber\\
P_n(\chi) &=& \frac{1}{2^n n!} \frac{d^n}{d\chi^n} 
\left(\chi^2 - 1\right)^n \, .
\ee

\begin{figure*}[htp]
  \begin{center}
   \includegraphics[type=pdf,ext=.pdf,read=.pdf,width=8cm]{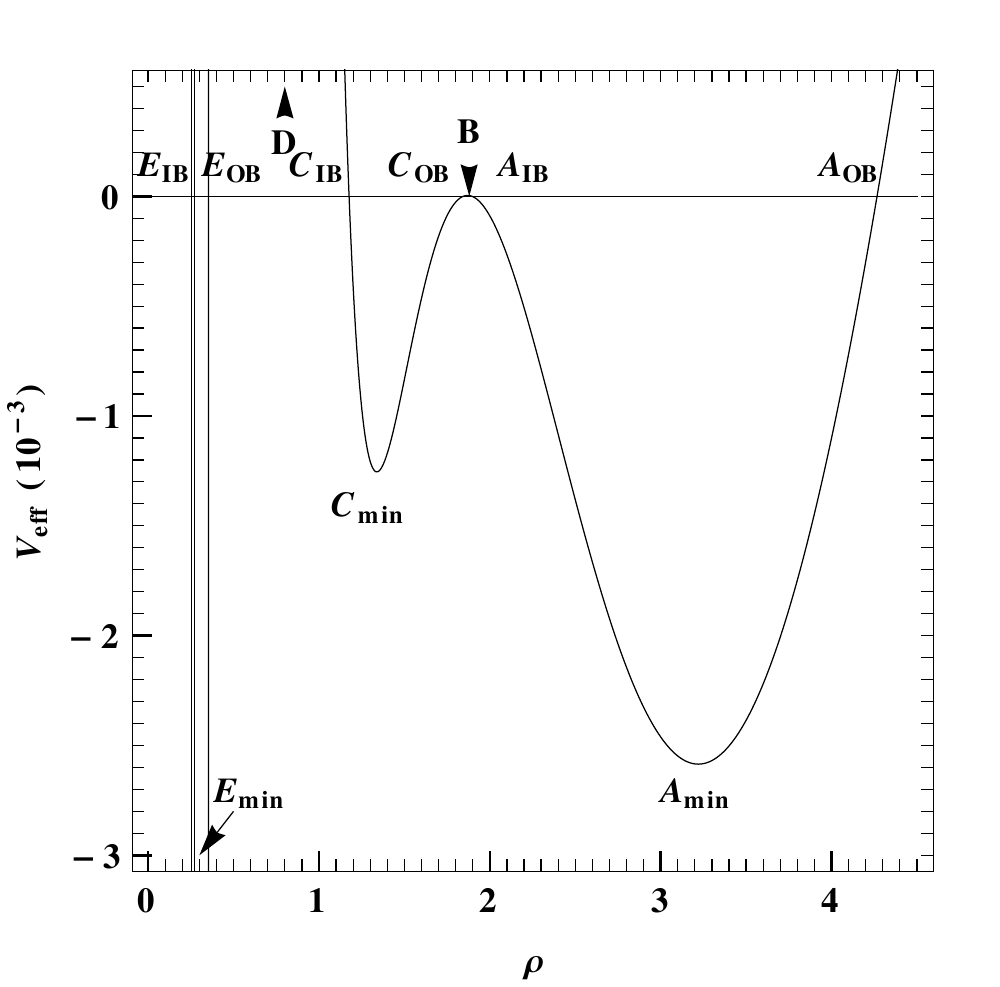}
   \includegraphics[type=pdf,ext=.pdf,read=.pdf,width=8cm]{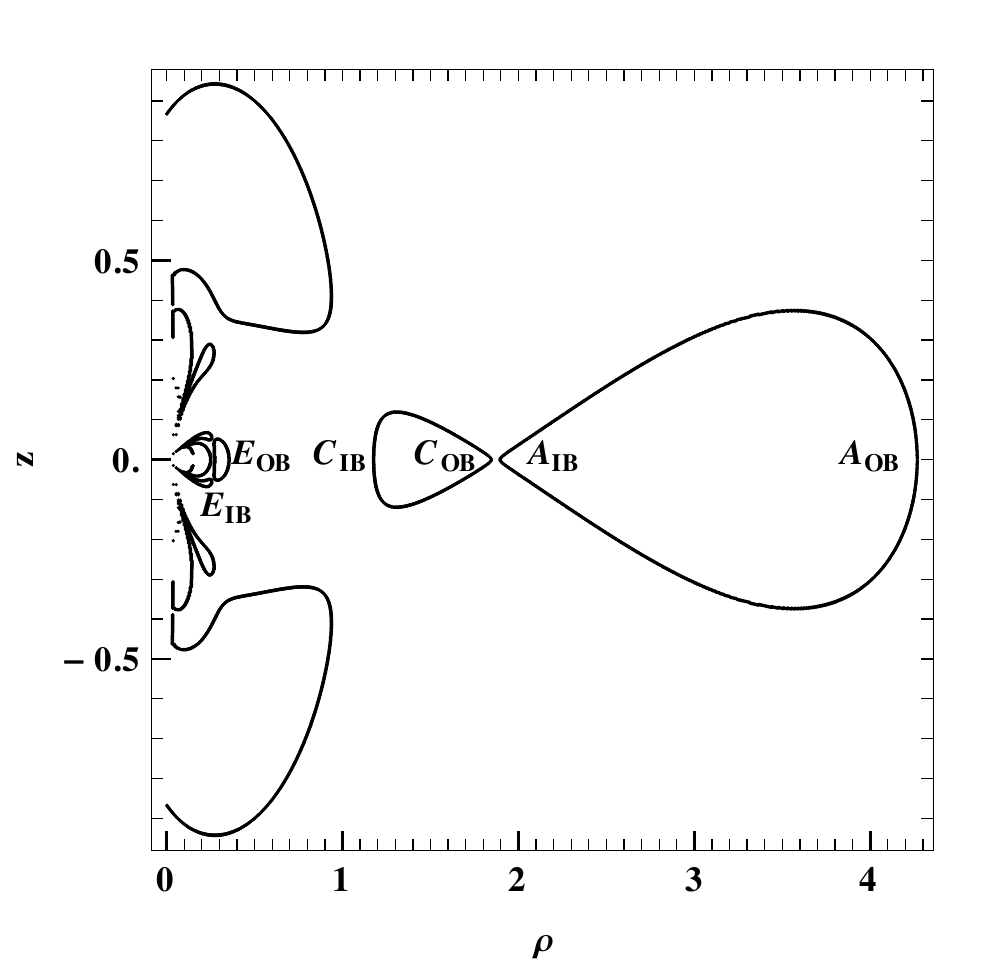}
 \caption{ The left panel shows the profile of the effective potential
 on the equatorial plane, $V_{\rm eff}(\rho,z=0)$, for $M=1$, $a_*=0.5$, $q=-0.18$,
 $h=-0.024$, $E=0.9143$, and $L_z=2.83$. $V_{\rm eff}=0$ (black horizontal line)
 is the limit of the allowed region. The roots of the effective potential, i.e.,
 $A_{OB}$, $A_{IB}$, $C_{OB}$, $C_{IB}$, $E_{OB}$, and $E_{IB}$, determine the
 limits of three disconnected regions of bounded non-plunging orbits. The index OB
 stands for outer border, while IB for inner border. The minima $A_{\rm min}$,
 $C_{\rm min}$, and $E_{\rm min}$ of $V_{\rm eff}$ indicate the existence of 
 stable circular orbits, while the maxima $B$ and $D$ indicate the existence 
 of unstable circular orbits. The right panel shows the limits of the allowed region
 $V_{\rm eff}=0$ on the $\rho-z$ plane, known as CZVs. }
\label{FigVeff}
  \end{center}
\end{figure*}

The MN solutions have an infinite number of free parameters: $\alpha$, which
regulates the spin $J$ of the central compact object; $k$, which regulates 
the  mass $M$ of the central compact object; and $\alpha_n$
($n=1, . . . , +\infty$) which regulate the mass-multipole moments, starting from 
the dipole $\alpha_1$, to the quadrupole  $\alpha_2$, etc. For $\alpha \neq 0$ 
and $\alpha_n = 0$, the metric reduces to the Kerr solution. For $\alpha = 
\alpha_n = 0$, it reduces to the Schwarzschild one. For $\alpha = 0$ and 
$\alpha_n \neq 0$, we obtain the static Weyl metric.

Without loss of generality, we can put $\alpha_1 = 0$ to bring the massive 
object to the origin of the coordinate system. In what follows, we restrict our 
attention to the subclass of MN spacetimes with $\alpha_n = 0$ for $n \neq 2$
and 4\footnote{A self-gravitating fluid is expected to be symmetric with respect
to the equatorial plane. In this case, all the $\alpha_n$ with $n$ odd must vanish
and the $n=2$ and 4 are the two possible independent leading order corrections 
to the Kerr geometry of the MN metrics (in the MN solutions, the current moments 
are not independent; e.g. the current-octupole moment is set by the mass-quadrupole moment).}. 
We then have four free parameters ($k$, $\alpha$, $\alpha_2$, $\alpha_4$) 
related to the mass $M$, the dimensionless spin parameter $a_* = J/M^2$, 
the anomalous mass-quadrupole moment $q = -(Q - Q_{\rm Kerr})/M^3$, and the
anomalous mass-hexadecapole moment $h = -(H - H_{\rm q})/M^5$, where $H_{\rm q}$
is the mass-hexadecapole moment of the object with $h=0$ and for $q=0$ it
reduces to $H_{\rm Kerr} = J^4/M^3$, by the relations
\be
\alpha &=& \frac{\sqrt{1 - a^2_*} - 1}{a_*} \, , \\
k &=& M \frac{1 - \alpha^2}{1 + \alpha^2} \, , \\
\alpha_2 &=& q \frac{M^3}{k^3} \, , \\
\alpha_4 &=& h \frac{M^5}{k^5} \, .
\ee
Note that $q$ measures the deviation from the quadrupole moment of a Kerr BH, 
but when $q \neq 0$ also the higher-order mass-multipole moments have a 
different value than in Kerr. For this reason, $h$ measures the deviation from the
hexadecapole moment of a Kerr BH only when $q = 0$.

The relation between prolate spheroidal and quasi-cylindrical coordinates is given
by
\be
\rho = k \sqrt{\left(x^2 - 1\right)\left(1 - y^2\right)} \, ,
\qquad
z = kxy \, ,
\ee
with inverse
\begin{align}
&x = \frac{1}{2k} \left(\sqrt{\rho^2 + \left(z + k\right)^2}
+ \sqrt{\rho^2 + \left(z - k\right)^2}\right) \, , \nonumber\\
&y = \frac{1}{2k} \left(\sqrt{\rho^2 + \left(z + k\right)^2}
- \sqrt{\rho^2 + \left(z - k\right)^2}\right) \, .
\end{align}
The standard Boyer-Lindquist coordinates $(r,\theta)$ are related to the 
prolate spheroidal coordinates and the quasi-cylindrical coordinates by
\be
&& r = k x + M \, , 
\nonumber\\
&& \cos\theta = y \, ,
\ee
and
\be
\rho &=& \sqrt{r^2 - 2 M r + a^2_* M^2} \sin\theta \, , 
\nonumber\\
z &=& (r - M) \cos\theta \, .
\ee

Because the MN metrics are stationary and axisymmetric, geodesic orbits have 
two constants of motion, the specific energy  $E = -u_t$ and the $z$-component 
of the specific angular momentum $L_z = u_\phi$. The $t$- and $\phi$-components 
of the 4-velocity of a test-particle are therefore 
\be
u^t &=& \dot{t} =\frac{E g_{\phi\phi} + L_z g_{t\phi}}{
g_{t\phi}^2 - g_{tt} g_{\phi\phi}} \, , \\ 
u^\phi &=& \dot{\phi} =- \frac{E g_{t\phi} + L_z g_{tt}}{
g_{t\phi}^2 - g_{tt} g_{\phi\phi}} \, .
\ee
where the dots denote derivatives with respect to the proper time. From the 
normalization of the 4-velocity $g_{\mu\nu} u^\mu u^\nu = -1$, we can write
\be
\dot{\rho}^2+\dot{z}^2 +V_{\rm eff}(E,L,\rho,z)=0\, ,
\ee
where the effective potential $V_{\rm eff}$ is defined by
\be
\label{eq-Veff}
V_{\rm eff} =\frac{f}{e^{2\gamma}}\left[1-\frac{E^2}{f}
+ \frac{f}{\rho^2}\left(L_z - \omega E\right)^2 \right] \, .
\ee
Due to stationarity and the axisymmetricity, the study of the geodesic
dynamics can be restricted on a meridian plane $(\phi = \mathrm{const})$, which
is implied by Eq.~(\ref{eq-Veff}) as well. Moreover, the motion takes place inside a 
region defined by $V_\mathrm{eff} \le 0$ (left panel of Fig.~\ref{FigVeff}), whose
border is the curve of zero velocity (CZV) $V_\mathrm{eff} = 0$
(right panel of Fig.~\ref{FigVeff}).

Circular orbits on the equatorial plane must have  $\dot{\rho} = \dot{z} = 0$,
which  implies $V_{\rm eff} = 0$, and $\ddot{\rho} = \ddot{z} = 0$, which implies
$\partial_\rho V_{\rm eff} = 0$ and $\partial_z V_{\rm eff} = 0$. This means that
circular equatorial orbits are located at simultaneous zeros and extrema of the
effective potential. The extrema shown in the left panel of Fig. \ref{FigVeff}
indicate the existence of these circular orbits, the minima $A_{\rm min}$,
$C_{\rm min}$, and $E_{\rm min}$ indicate the existence of stable circular 
orbits, while the maxima $B$ and $D$ indicate the existence of unstable 
circular orbits. In order to calculate the positions of the circular
orbits, we must also demand that the effective potential to be equal to zero. Thus,
because $\partial_z V_{\rm eff} = 0$ is satisfied identically for $z=0$ (simply
because of the reflection symmetry of our MN metric with respect to the equatorial
plane), from the aforementioned conditions one can obtain $E$ and $L_z$ as a
function of the radius of the orbit $\rho$:
\be
E &=& - \frac{g_{tt} + g_{t\phi}\Omega}{
\sqrt{-g_{tt} - 2g_{t\phi}\Omega - g_{\phi\phi}\Omega^2}} \, , \\
L_z &=& \frac{g_{t\phi} + g_{\phi\phi}\Omega}{
\sqrt{-g_{tt} - 2g_{t\phi}\Omega - g_{\phi\phi}\Omega^2}} \, ,
\ee
where
\be 
\Omega = 
\frac{- \partial_r g_{t\phi} 
\pm \sqrt{\left(\partial_r g_{t\phi}\right)^2 
- \left(\partial_r g_{tt}\right) \left(\partial_r 
g_{\phi\phi}\right)}}{\partial_r g_{\phi\phi}}
\ee
is the angular frequency and the sign $+$ $(-)$ is for orbits with angular momentum
parallel (antiparallel) to the spin of the compact object. These orbits are stable 
under small perturbations in the radial direction if $\partial_\rho^2 V_{\rm eff} > 0$, 
and in the vertical direction if $\partial_z^2 V_{\rm eff} > 0$. 

\begin{figure}[htp]
  \begin{center}
  \includegraphics[type=pdf,ext=.pdf,read=.pdf,width=8cm]{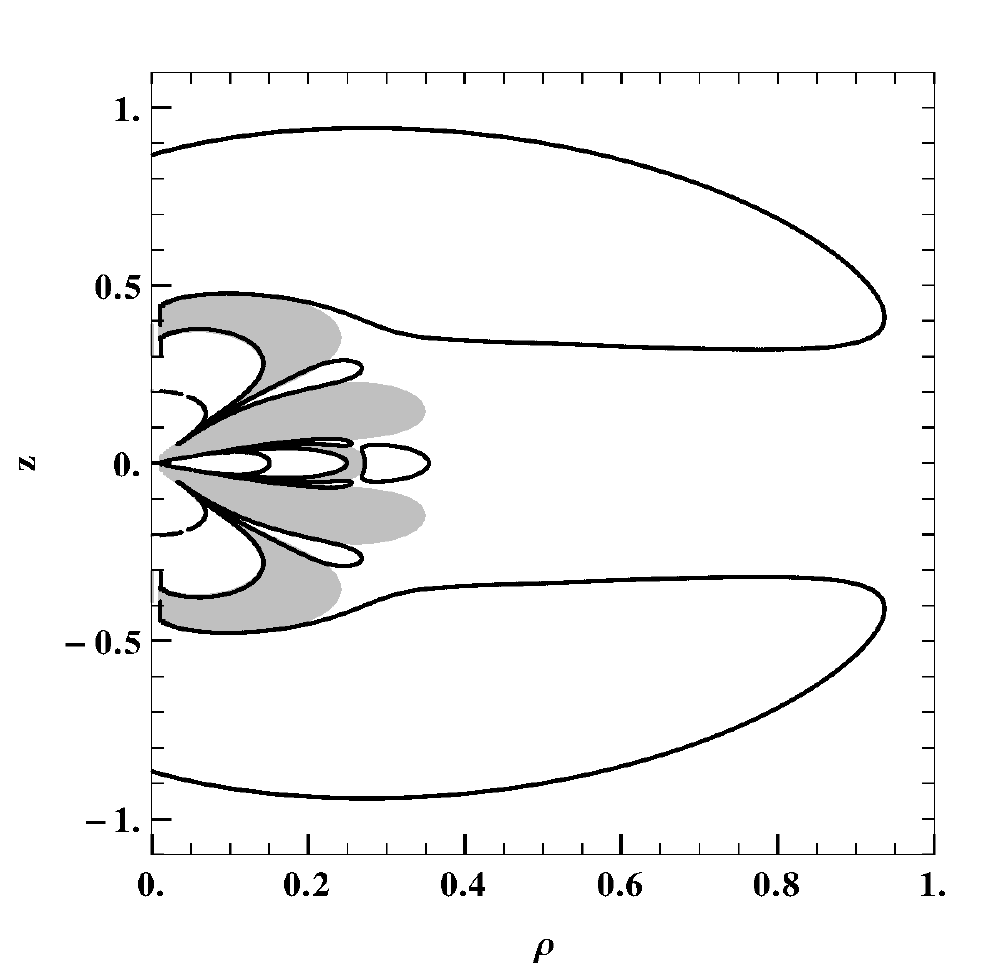}
  \end{center}
 \caption{The region with CTCs (in gray) is determined by the condition $g_{\phi\phi}<0$, 
 while the regions where motion is permitted are delimited by the CZV (black curves). 
 It seems that the two regions never overlap for plausible values of the orbital parameters. 
 The MN parameters, $E$, and $L_z$ are the same as in Fig.~\ref{FigVeff}.  }
\label{FigCTC}

\end{figure}

The no-hair theorem~\cite{hair} states that the only asymptotically flat, vacuum 
and stationary solution of the Einstein's equations that is non-singular on and 
outside an event horizon and that presents no closed time-like curves (CTCs)
outside it is given by the Kerr metric. Therefore, the MN spacetimes must either
have no event horizon, or present naked singularities or CTCs outside it. In fact,
a MN metric can present the last two anomalies. Namely, the surface $x = 1$
($\rho=0$, $|z|\leq k$), which is the event horizon in the Kerr case $\alpha_n = 0$,
is in general ($\alpha_n \ne 0$) only a partial horizon, because it presents
a naked curvature singularity on the equatorial plane (i.e. at $x=1$, $y=0$,
corresponding to $\rho=z=0$). Moreover, there are CTCs outside it (Fig.~\ref{FigCTC}). 
However, it seems that the CTCs and the allowed region $V_{\rm eff} < 0$ 
do not overlap.

Nevertheless, in our scenario the exotic object is compact enough so that the 
plunging particles can reach the inner stable orbits, but not so much to have 
the region with CTCs. Furthermore, the pathological region with CTCs can 
even be removed by exciting the spacetime. This occurs, for example, if the 
spacetime goes to a new phase and a domain wall is formed~\cite{exotic1,exotic2}. 
Across the domain wall, the metric is nondifferentiable and the expected region 
with closed timelike curves arises from the naive continuation of the metric 
ignoring the domain wall. The latter can be made of very exotic stuff, and 
in the literature there are examples of supertubes~\cite{exotic1} and fundamental 
strings~\cite{exotic2}.

\section{Properties of equatorial circular orbits in MN spacetimes}

\subsection{MN spacetimes with $q \neq 0$ and $h = 0$ \label{ss-1}}

This is the case already discussed in the literature. The basic properties
of equatorial circular orbits can be understood by considering a specific
spin parameter $a_*$ and by changing the anomalous quadrupole moment 
$q$. As an example, we report the values for the case $a_* = 0.5$ (see Tab.~\ref{tab}).

For $q=0$, we recover the Kerr spacetime. Outside the even horizon, the 
spacetime is everywhere regular (there are no CTCs). 
Circular orbits on the equatorial plane exist for radii $\rho > \rho_{\rm PO}$,
they are always vertically stable, but they are radially stable only for 
$\rho > \rho_{\rm ISCO}$. If the compact object is surrounded by a thin 
accretion disk, the inner edge of the disk is at $\rho_{\rm ISCO}$.
For $a_* = 0.5$, we have $\rho_{\rm PO} = 1.03$ and $\rho_{\rm ISCO} = 3.11$
(here and in the following, we use units in which $M=1$).

MN spacetimes with $q \neq 0$ have regions with CTCs just outside the 
event horizon. That is true even on the equatorial plane. When the compact 
object is more oblate than a Kerr BH ($q > 0$), there are no interesting new 
properties with respect to the Kerr metric. As the object becomes more and 
more oblate ($q$ increases), the value of $\rho_{\rm ISCO}$ increases. For the
case $q = 0.20$ shown in Tab.~\ref{tab}, equatorial circular orbits exist for
$\rho > 1.39$, they are always vertically stable, 
but they are radially stable only for $\rho > \rho_{\rm ISCO} = 3.50$. Unlike the 
Kerr background, a narrow region with circular orbits appears even at
very small radii ($\rho \approx 0.04$ for $q = 0.20$). 
This narrow region extends up to the region with CTCs. Even if it
seems that this region never overlaps with the CTC region,
it is more likely unphysical: these orbits have negative energy, suggesting that
their possible existence would make the object 
unstable~\cite{ergoregion}. While orbits with 
negative energy exist even in the ergoregion of a Kerr BH, they are never 
stable.

\begin{figure}
\begin{center}  
\includegraphics[type=pdf,ext=.pdf,read=.pdf,width=8cm]{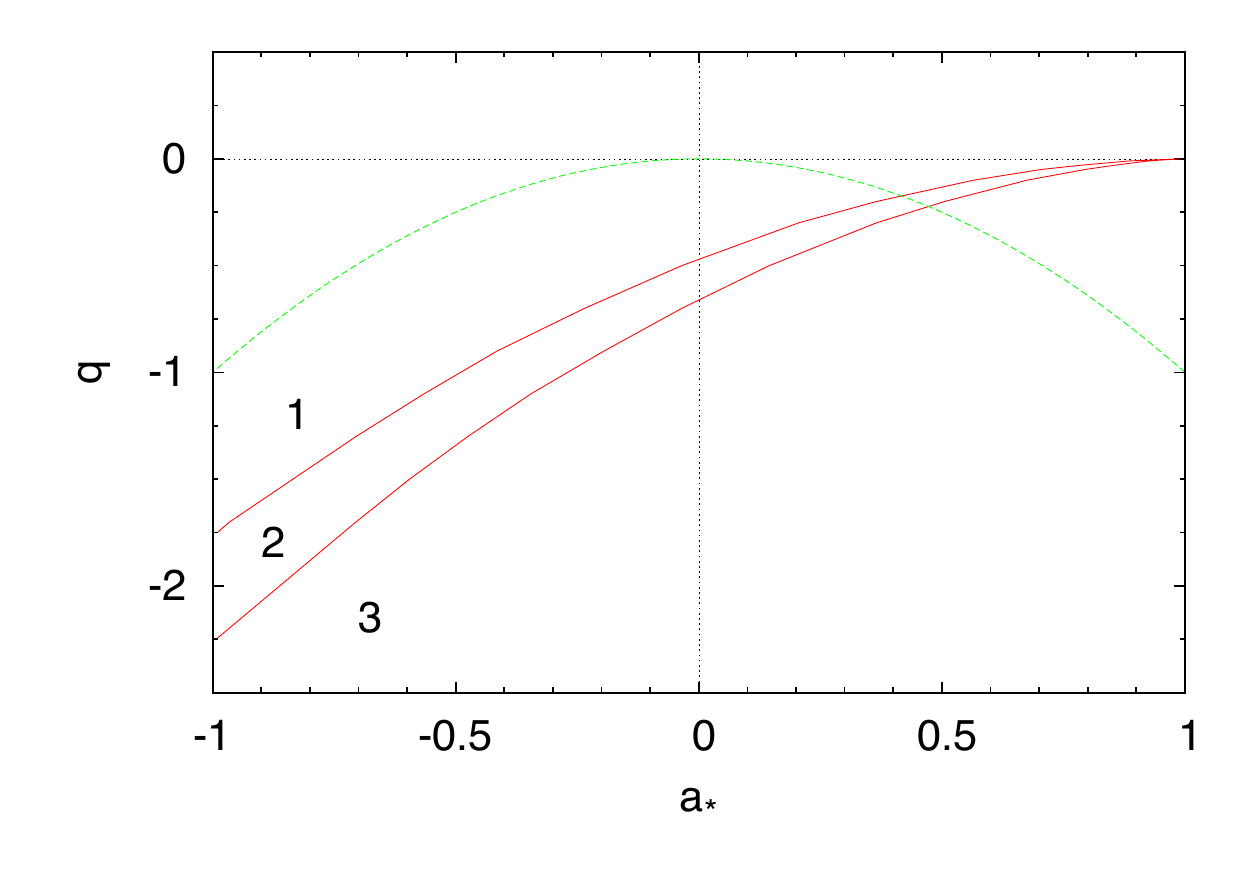}
 \end{center}
 \vspace{-0.5cm}
  \caption{MN spacetimes with non-vanishing anomalous quadrupole moment $q$. The
   solid red curves separate spacetimes with qualitatively different properties.
   Region 1: spacetimes in which equatorial circular orbits of astrophysical
   interest are located at $\rho > \rho_{\rm ISCO}$ and the orbit at
   $\rho = \rho_{\rm ISCO}$ is marginally stable along the radial direction.
   Region 2: spacetimes in which equatorial circular orbits of astrophysical
   interest are located at $\rho > \rho_{\rm ISCO}$ and at 
   $\rho_1 < \rho < \rho_3$ ($\rho_3 <  \rho_{\rm ISCO}$);
   the orbit at $\rho = \rho_1$ is marginally stable along the vertical direction,
   while the one at $\rho = \rho_{\rm ISCO}$ is marginally stable along the radial
   direction; the energy of the orbits at $\rho = \rho_3$ and at 
   $\rho = \rho_{\rm ISCO}$ is the same and therefore the gas's particles can plunge
   from the inner edge of the external disk to the outer edge of the internal disk.
   Region 3: spacetimes in which equatorial circular orbits of astrophysical
   interest are located at $\rho > \rho_1$ and the orbit at $\rho = \rho_1$ is
   marginally stable along the vertical direction. The compact object is oblate if
   $q > - a_*^2$; the green dashed curve separates oblate objects from prolate
   objects. The compact object is more oblate than a Kerr black hole if $q > 0$.
   See the text for details.}
\label{f-q}
\end{figure}

For $q<0$, the spacetime can have properties absent in the Kerr background and
that are potentially of astrophysical interest. 
Let us consider again the example with $a_*=0.5$.
For small $|q|$, there is a region with circular orbits at small radii. 
Such orbits are radially stable, but vertically unstable. 
If we consider the case $q = -0.01$ of Tab.~\ref{tab}, such a region is 
at $0.25 < \rho < 0.35$. At larger radii, circular orbits exist for $\rho > 0.99$, 
they are always vertically stable, but they are radially stable 
only for $\rho > \rho_{\rm ISCO} = 3.09$.
As $|q|$ increases, the size of the region with circular orbits at small radii 
increases as well, and eventually it merges with the one at larger radii. 
At this point, we have two disconnected regions with stable orbits,
$\rho_1 < \rho < \rho_2$ and $\rho > \rho_{\rm ISCO}$; this is the case of
$q = -0.1$, with $\rho_1 = 1.00$, $\rho_2 = 1.16$ and $\rho_{\rm ISCO} = 2.83$,
and of $q = -0.2$, with $\rho_1 = 1.24$, $\rho_2 = 2.00$ and $\rho_{\rm ISCO} = 2.17$. 
The orbit with radius $\rho = \rho_1$ is marginally unstable along the
vertical direction, while the one with $\rho = \rho_2$ is marginally unstable along
the radial direction. If the orbits in the region $\rho_1 < \rho < \rho_2$ have
 energy larger than the orbit at $\rho = \rho_{\rm ISCO}$ (the case $q = - 0.1$),
the region $\rho_1 < \rho < \rho_2$ is not of astrophysical interest, as
particles/gas orbiting around the compact object after reaching the radius
$\rho_{\rm ISCO}$ plunge onto the compact object, without populating the region
$\rho_1 < \rho < \rho_2$. However, as $|q|$ increases, the size of the region
$\rho_1< \rho < \rho_2$ also increases and there may exist a region 
$\rho_1< \rho < \rho_3$ with $\rho_3 < \rho_2$ in which the orbital energy is
smaller than the one at $\rho = \rho_{\rm ISCO}$ (the case $q = -0.2$). These
orbits are of astrophysical interest because they can be filled up by the gas of
accretion reaching the radius $\rho = \rho_{\rm ISCO}$. If $|q|$ continues
increasing, the region $\rho_2 < \rho < \rho_{\rm ISCO}$ with radially unstable
orbits disappears: now circular orbits are always radially stable, but they are
vertically stable for $\rho > \rho_1$, as is for the case $q = -0.3$, with $\rho_1 = 1.43$.

Fig.~\ref{f-q} shows the three different
kinds of MN spacetimes with $q \neq 0$ on the plane spin parameter--anomalous
quadrupole moment. The region 1, which includes the Kerr solution $q = 0$, has
equatorial stable circular orbits for $\rho > \rho_{\rm ISCO}$ and the orbit at
$\rho = \rho_{\rm ISCO}$ is marginally radially stable; for $q < 0$, these
spacetimes may also have equatorial stable circular orbits at smaller radii, but
they are not of astrophysical interest because these orbits require  energies higher than
the one of the orbit at $\rho = \rho_{\rm ISCO}$. In the region 2, there are
spacetimes with equatorial stable circular orbits of astrophysical interest for
 $\rho_1< \rho < \rho_3$ and $\rho > \rho_{\rm ISCO}$
 ($\rho_3 <  \rho_{\rm ISCO}$); the orbit at $\rho = \rho_1$
is marginally vertically stable, while the one at $\rho = \rho_{\rm ISCO}$ is
marginally radially stable. Lastly, the region 3 includes those spacetimes in which
equatorial circular orbits are always radially stable, but they are vertically 
stable only for $\rho > \rho_1$.

\begin{table*}
\begin{center}
\begin{tabular}{c c c c c c c c c c c c c}
\hline
\hline
$a_*$ &  \hspace{.2cm} & $q$ &  \hspace{.2cm} & $h$ &  \hspace{.2cm} & 
Circular orbits &  \hspace{.2cm} & $\partial_\rho^2 V_{\rm eff} > 0$ &
 \hspace{.2cm} & $\partial_z^2 V_{\rm eff} > 0$ &  \hspace{.2cm} & CTCs \\
\hline
\hline
0.5 && 0 && 0 && $\rho > 1.03$ && $\rho > 3.11$ && $\rho > 1.03$ && no CTCs \\
\hline
\hline
0.5 && 0.20 && 0 && $\rho \approx 0.04$ && $\rho \approx 0.04$ &&  && $\rho < 0.04$ \\
 &&  &&  && $\rho > 1.39$ && $\rho > 3.50$ && $\rho > 1.39$ &&  \\
\hline
0.5 && $-0.01$ && 0 && $0.25 < \rho < 0.35$ && $0.25 < \rho < 0.35$ &&  && $\rho < 0.15$ \\
 &&  &&  && $\rho > 0.99$ && $\rho > 3.09$ && $\rho > 0.99$ &&  \\
\hline
0.5 && $-0.10$ && 0 && $\rho > 0.56$ && $0.56 < \rho < 1.16$, $\rho > 2.83$ && $\rho > 1.00$ && $\rho < 0.35$ \\
\hline
0.5 && $-0.20$ && 0 && $\rho > 0.73$ && $0.73 < \rho < 2.00$, $\rho > 2.17$ && $\rho > 1.24$ && $\rho < 0.45$ \\
\hline
0.5 && $-0.30$ && 0 && $\rho > 0.85$ && $\rho > 0.85$ && $\rho > 1.43$ && $\rho < 0.52$ \\
\hline
\hline
\end{tabular}
\end{center}
\vspace{-0.2cm}
\caption{Regions with equatorial circular orbits (fourth column), radially stable orbits (fifth 
column), vertically stable orbits (sixth column), and CTCs (seventh column) in MN
spacetimes with spin parameter $a_* = 0.5$, anomalous mass-quadrupole moment $q$, 
and vanishing anomalous mass-hexadecapole moment $h$. See the text for details.}
\label{tab}
\end{table*}

\subsection{MN spacetimes with $q = 0$ and $h \neq 0$}

If the deformation parameter of the spacetime is the anomalous hexadecapole
moment $h$ instead of the $q$, the picture is qualitatively the same as the
one discussed in the previous subsection, with $h$ playing the role of $-q$
(the minus sign is just a matter of definition).

\subsection{MN spacetimes with $q < 0$ and $h \neq 0$}

Let us now discuss the more general scenario, when both $q$ and $h$ may be
non-vanishing, starting from the case with $q < 0$. As in Subsection~\ref{ss-1},
we can consider the specific case $a_* = 0.5$ (see Tab.~\ref{tab2}); 
for another value of $a_*$, the picture is qualitatively the same, but the same 
phenomena occur for different values of $q$ and $h$. First, we consider
a spacetime in which $|q|$ is large enough that for $h = 0$ we have stable
equatorial orbits for $\rho > \rho_1$. In Tab.~\ref{tab2}, we consider the specific 
case $q = -0.25$, where $\rho_1 = 1.34$ when $h = 0$. If we change the anomalous
hexadecapole moment to $h = -0.03$, we find that circular orbits exist even in two 
narrow region at small radii: a region close to the one with CTCs ($\rho \approx 0.27$), 
and another one in which the orbits are vertically stable but radially unstable
($0.40 < \rho < 0.43$). The latter expands as $|h|$ increases and it merges with 
the region of circular orbits at large radii. For instance, the regions $0.40 < \rho < 0.43$
and $\rho > 0.68$ for $h = -0.03$ become the region $\rho > 0.53$ when 
$h = -0.05$. At this point, we have three regions of stable circular orbits:
one close to the pathological space with CTCs, one at $\rho_4 < \rho < \rho_5$, 
and the last one for $\rho > \rho_1$. The orbit at $\rho = \rho_4$ is marginally 
radially stable, while the one at $\rho = \rho_5$ is marginally vertically stable. 
For $h = -0.05$, the region close to the one with CTCs is at $\rho \approx 0.32$, 
while $\rho_4 = 0.67$, $\rho_5 = 0.70$, and $\rho_1 = 1.17$. A spacetime with 
similar properties does not exist if the only non-vanishing deformation parameter is 
$q$. Let us notice that the region $\rho_4 < \rho <\rho_5$ may include a subregion 
of astrophysical interest $\rho_4 < \rho < \rho_6$, where $\rho_6 < \rho_5$, in
which the particles have energies smaller than the one of a particle at
$\rho = \rho_1$. In presence of an accretion disk, the region $\rho_4 < \rho < \rho_6$ 
could be filled by the accreting gas. For larger values of $|h|$, the gap 
$\rho_5 < \rho < \rho_1$ disappears and stable orbits exist for $\rho > \rho_4$.
This is the case with $h = -0.08$, where $\rho_4 = 0.91$. By continuing to increase the 
value of $|h|$, a new gap shows up, in which the orbits are vertically stable but 
radially unstable. So, for $h = -0.1$ stable orbits exist for $\rho_4 < \rho < \rho_7$ 
and $\rho > \rho_8$, where $\rho_4 = 1.14$, $\rho_7 = 1.71$, and $\rho_8 = 2.04$.
The new gap is $\rho_7 < \rho < \rho_8$ and we have used a different notation
with respect to the gap $\rho_5 < \rho < \rho_1$ of the case $h = -0.05$ because it
has different properties: $\rho_7$ and $\rho_8$ are marginally radially stable, 
while $\rho_5$ and $\rho_1$ were marginally vertically stable. If we continue
increasing the value of $|h|$, the gap $\rho_7 < \rho < \rho_8$ disappears.
This is the case $h = -0.15$: the spacetime now has features similar to that 
of a MN background with $q > 0$ and vanishing $h$, in which stable orbits exist 
for $\rho > \rho_4$ ($\rho_4 = 2.32$ for $h = -0.15$), the orbit at the radius 
$\rho = \rho_4$ is marginally radially stable, and the value of $\rho_4$ increases 
as $|h|$ increases.

If we start with $q = -0.25$ and $h = 0.0$ and we increase the value of $h$ (now $h > 0$), 
we simply move $\rho_1$ at larger radii. For example, for $h = 1.0$ we have 
$\rho_1 = 2.06$, while for $h = 5.0$ we have $\rho_1 = 2.83$. We never form
islands of stable circular orbits at small radii.

The properties of the MN spacetime may be even more interesting when $q$ is 
still negative, but $|q|$ is not so large, so that for $h = 0$ the region 
$\rho_1 < \rho < \rho_2$ is still separated by a gap from the one 
$\rho > \rho_{\rm ISCO}$. As a specific example, we have $q = -0.15$
in Tab.~\ref{tab2}. The properties of the equatorial circular orbits can be understood 
on the basis of the previous discussion. When $h = -0.01$, $\rho_1=1.08$, 
$\rho_2=1.40$, and $\rho_{\rm ISCO}=2.64$. For smaller $h$, we may
have three disconnected regions of stable circular orbits (and, at least in some 
cases, of astrophysical interest): a new region at $\rho_9 < \rho < \rho_{10}$, 
the regions $\rho_1 < \rho < \rho_2$, and the one at $\rho > \rho_{\rm ISCO}$.
In the specific case $h = -0.029$, we find $\rho_9 = 0.68$, 
$\rho_{10} = 0.77$, $\rho_1 = 0.89$, $\rho_2 = 1.22$, and $\rho_{\rm ISCO} = 2.67$.

\begin{table*}
\begin{center}
\begin{tabular}{c c c c c c c c c c c c c}
\hline
\hline
$a_*$ &  \hspace{.2cm} & $q$ &  \hspace{.2cm} & $h$ &  \hspace{.2cm} & 
Circular orbits &  \hspace{.2cm} & $\partial_\rho^2 V_{\rm eff} > 0$ &
 \hspace{.2cm} & $\partial_z^2 V_{\rm eff} > 0$ &  \hspace{.2cm} & CTCs \\
\hline
\hline
0.5 && $-0.25$ && 0.00 && $\rho > 0.79$ && $\rho > 0.79$ && $\rho > 1.34$ && $\rho < 0.49$ \\
\hline
0.5 && $-0.25$ && $-0.03$ && $\rho \approx 0.27$ && $\rho \approx 0.27$ && $\rho \approx 0.27$ && $\rho < 0.27$ \\
 &&  &&  && $0.40 < \rho < 0.43$ && && $0.40 < \rho < 0.43$ &&  \\
 &&  &&  && $\rho > 0.68$ && $\rho > 0.68$ && $\rho > 1.26$ &&  \\
\hline
0.5 && $-0.25$ && $-0.05$ && $\rho \approx 0.32$ && $\rho \approx 0.32$ && $\rho \approx 0.32$ && $\rho < 0.32$ \\
&&  &&  && $\rho > 0.53$ && $\rho > 0.67$ && $0.53 < \rho < 0.70$, $\rho > 1.17$ &&  \\
\hline
0.5 && $-0.25$ && $-0.08$ && $\rho \approx 0.38$ && $\rho \approx 0.38$ &&  && $\rho < 0.38$ \\
 &&  &&  && $\rho > 0.69$ && $\rho > 0.91$ && $\rho > 0.69$ &&  \\
\hline
0.5 && $-0.25$ && $-0.10$ && $\rho \approx 0.41$ && $\rho \approx 0.41$ && $\rho \approx 0.41$ && $\rho < 0.41$ \\
 &&  &&  && $\rho > 0.77$ && $1.14 < \rho < 1.71$, $\rho > 2.04$ && $\rho > 0.77$ &&  \\
\hline
0.5 && $-0.25$ && $-0.15$ && $\rho \approx 0.45$ && $\rho \approx 0.45$ &&  && $\rho < 0.45$ \\
 &&  &&  && $\rho > 0.92$ && $\rho > 2.32$ && $\rho > 0.92$ &&  \\
\hline
0.5 && $-0.25$ && 1.00 && $\rho > 1.35$ && $\rho > 1.35$ && $\rho > 2.06$ && $\rho < 0.87$ \\
\hline
0.5 && $-0.25$ && 5.00 && $\rho > 1.88$ && $\rho > 1.88$ && $\rho > 2.83$ && $\rho < 1.19$ \\
\hline
\hline
0.5 && $-0.15$ && $-0.010$ && $\rho \approx 0.21$ && $\rho \approx 0.21$ &&  && $\rho < 0.21$ \\
 &&  &&  && $0.30 < \rho < 0.31$ &&  && $0.30 < \rho < 0.31$ &&  \\
 &&  &&  && $\rho > 0.59$ && $0.59< \rho < 1.40$, $\rho > 2.64$ && $\rho > 1.08$ &&  \\
\hline
0.5 && $-0.15$ && $-0.029$ && $\rho \approx 0.30$ && $\rho \approx 0.30$ &&  && $\rho < 0.30$ \\
 &&  &&  && $\rho > 0.54$ && $0.68 < \rho < 1.22$, $\rho > 2.67$ && $0.54 < \rho < 0.77$, $\rho > 0.89$ &&  \\
\hline
0.5 && $-0.15$ && $-0.035$ && $\rho \approx 0.32$ && $\rho \approx 0.32$ &&  && $\rho < 0.32$ \\
 &&  &&  && $\rho > 0.61$ && $0.80 < \rho < 1.13$, $\rho > 2.67$ && $\rho > 0.61$ &&  \\
\hline
0.5 && $-0.15$ && $-0.050$ && $\rho \approx 0.36$ && $\rho \approx 0.36$ &&  && $\rho < 0.36$ \\
 &&  &&  && $\rho > 0.74$ && $\rho > 2.69$ && $\rho > 0.74$ &&  \\
\hline
\hline
\end{tabular}
\end{center}
\vspace{-0.2cm}
\caption{Regions with equatorial circular orbits (fourth column), radially stable orbits (fifth column),
vertically stable orbits (sixth column), and CTCs (seventh column) in MN
spacetimes with spin parameter $a_* = 0.5$, anomalous mass-quadrupole moment $q < 0$, 
and anomalous mass-hexadecapole moment $h$. See the text for details.}
\label{tab2}
\end{table*}

\subsection{MN spacetimes with $q > 0$ and $h \neq 0$}
 
The properties of the case $q > 0$ can be easily expected on the basis of 
the results found in the previous subsections. The effect of increasing/decreasing 
$h$ is equivalent to that of decreasing/increasing $q$.
Tab.~\ref{tab3} shows the case $q = 0.25$.

\subsection{Islands of stable equatorial circular orbits of astrophysical interest}

The study of the previous subsections suggests the following conjecture:
the MN spacetimes with arbitrary deformation parameters ($q$, $h$, and higher
order deformations) can potentially have several disconnected regions
with stable equatorial orbits of astrophysical interest. Accreting compact objects
described by the MN solutions, if they exist, might thus be surrounded by 
Saturn-like structures, in which the accreting gas can fill different rings around
the compact object. In the case $a_* = 0.5$, $q = -0.15$ and $h = -0.029$, 
we can have an accretion disk with two internal rings. However, for non-vanishing 
higher order deformations, three or more rings may be possible.

\begin{table*}
\begin{center}
\begin{tabular}{c c c c c c c c c c c c c}
\hline
\hline
$a_*$ &  \hspace{.2cm} & $q$ &  \hspace{.2cm} & $h$ &  \hspace{.2cm} & 
Circular orbits &  \hspace{.2cm} & $\partial_\rho^2 V_{\rm eff} > 0$ &
 \hspace{.2cm} & $\partial_z^2 V_{\rm eff} > 0$ &  \hspace{.2cm} & CTCs \\
\hline
\hline
0.5 && 0.25 && 0.0 && $\rho \approx 0.42$ && $\rho \approx 0.42$ &&  && $\rho < 0.42$ \\
 &&  &&  && $\rho > 1.44$ && $\rho > 3.58$ && $\rho > 1.44$ &&  \\
\hline
0.5 && 0.25 && $-1.0$ && $\rho \approx 0.77$ && $\rho \approx 0.77$ && $\rho \approx 0.77$ && $\rho < 0.77$ \\
 &&  &&  && $\rho > 1.75$ && $\rho > 3.79$ && $\rho > 1.75$ &&  \\
\hline
0.5 && 0.25 && 0.1 && $0.59 < \rho < 0.67$ && $0.59 < \rho < 0.67$ &&  && $\rho < 0.42$ \\
 &&  &&  && $\rho > 1.37$ && $\rho > 3.55$ && $\rho > 1.37$ &&  \\
\hline
0.5 && 0.25 && 0.5 && $\rho > 0.96$ && $0.96 < \rho < 1.55$, $\rho > 3.43$ && $\rho > 1.43$ && $\rho < 0.67$ \\
\hline
0.5 && 0.25 && 1.0 && $\rho > 1.17$ && $1.17 < \rho < 2.13$, $\rho > 3.21$ && $\rho > 1.73$ && $\rho < 0.79$ \\
\hline
0.5 && 0.25 && 2.0 && $\rho > 1.40$ && $\rho > 1.40$ && $\rho > 2.07$ && $\rho < 0.93$ \\
\hline
\hline
\end{tabular}
\end{center}
\vspace{-0.2cm}
\caption{Regions with equatorial circular orbits (fourth column), radially stable orbits (fifth column),
vertically stable orbits (sixth column), and CTCs (seventh column) in MN spacetimes 
with spin parameter $a_* = 0.5$, anomalous mass-quadrupole moment $q>0$, 
and anomalous mass-hexadecapole moment $h$. See the text for details.}
\label{tab3}
\end{table*}

\section{Observational signature}

Let us consider a spacetime in which equatorial circular orbits are stable for
$\rho > \rho_{\rm ISCO}$ and in the inner regions $\rho_1 < \rho < \rho_2$,
$\rho_4 < \rho < \rho_5$ , etc. at smaller radii. In the presence of a gas of accretion,
we may have the usual thin disk at $\rho > \rho_{\rm ISCO}$ and several gas rings for 
$\rho_1 < \rho < \rho_3$, $\rho_4 < \rho < \rho_6$, etc. Is the thermal spectrum of
this multi-part accretion disk very different from the one we can expect around a
Kerr BH with inner edge $\rho_{\rm ISCO}'$? The answer is no and the reason is 
the following. The high-energy part of the
thermal spectrum of a thin accretion  disk around a compact object depends on the
radiative efficiency $\eta = 1 - E_{\rm in}$, where $E_{\rm in}$ is the specific
energy of the gas particles at the inner edge of the disk. So, it does not matter
if the disk has some gaps. The thermal spectrum of our disk will look like the one
around a Kerr BH with the same radiative  efficiency. 
As the gas in inner rings necessarily has energy lower than the one
at $\rho_{\rm ISCO}$, the radiative efficiency of these objects is higher than
$\eta = 1 - E_{\rm ISCO}$, which might be (erroneously) interpreted as a higher
value of the spin parameter of the object if we do not take the radiation from 
the inner disk into account.
The ring structure of the accretion disk may be observed in high-resolution 
images of the disk itself, see e.g. Figs.~5 and 6 of Ref.~\cite{apj12}, but this is 
definitively out of reach for present and near future facilities.

How can we then identify the existence of islands of stable
orbits around a BH candidate? We think that a clear observational signature of
these non-Kerr spacetimes may be found in the gravitational wave signal emitted by
an extreme-mass ratio inspiral (EMRI) system, in which a stellar-mass compact
object orbits around a super-massive BH candidate. In the standard case, the
frequency and the amplitude of the waveform of the gravitational radiation are 
expected to increase regularly as the orbit of the small body shrinks to the ISCO 
radius -- this is usually called the ``chirp'' phase in the literature. At
the ISCO, the orbit becomes radially unstable and the stellar-mass body plunges
into the super-massive BH candidate within a dynamical time-scale, emitting an
irregular waveform. In the presence of inner regions of stable
orbits, the picture may change as follows. As in the Kerr metric, the frequency and
the amplitude of the waveform of the gravitational radiation should increase regularly
as the small body approaches the ISCO radius. However, once at the ISCO, the small
body may not plunge into the central object, but rather land in an inner region with stable
orbits. During the transition from the ISCO to the island, the
system would emit some complicated waveform of gravitational waves to adjust the
small body to the new orbit, but, after that, we should start observing again the
typical EMRI waveform, in which the frequency and the amplitude increase regularly.
This process may be repeated several times, with the small body that reaches the
inner edge of an outer region and jumps to another region at smaller radii. The
whole waveform should thus consist of several pieces of regular waveforms (the
usual EMRI-type signal) broken by ``bursts'', due to gravitational waves
emitted when the small body adjusts into the new orbit. 

\begin{figure*}[htp]
  \begin{center}
  \includegraphics[type=pdf,ext=.pdf,read=.pdf,width=8cm]{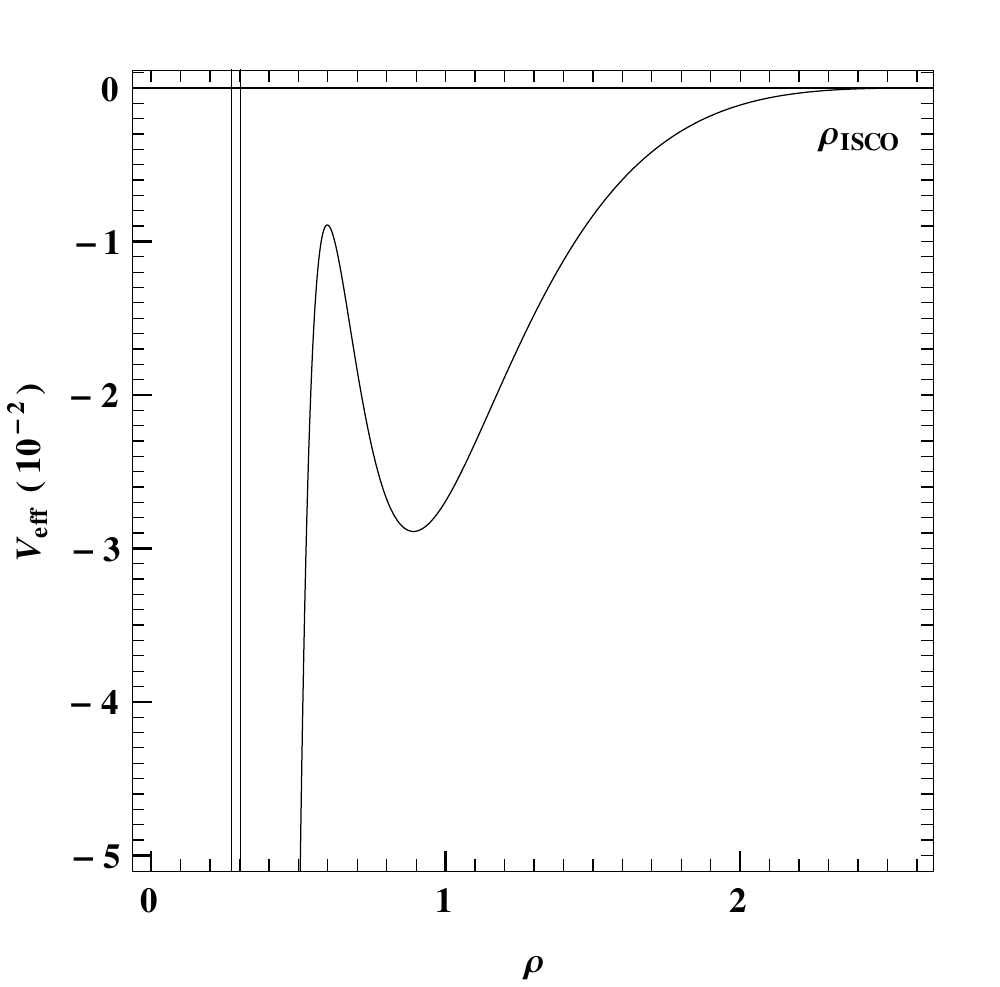}
  \includegraphics[type=pdf,ext=.pdf,read=.pdf,width=8cm]{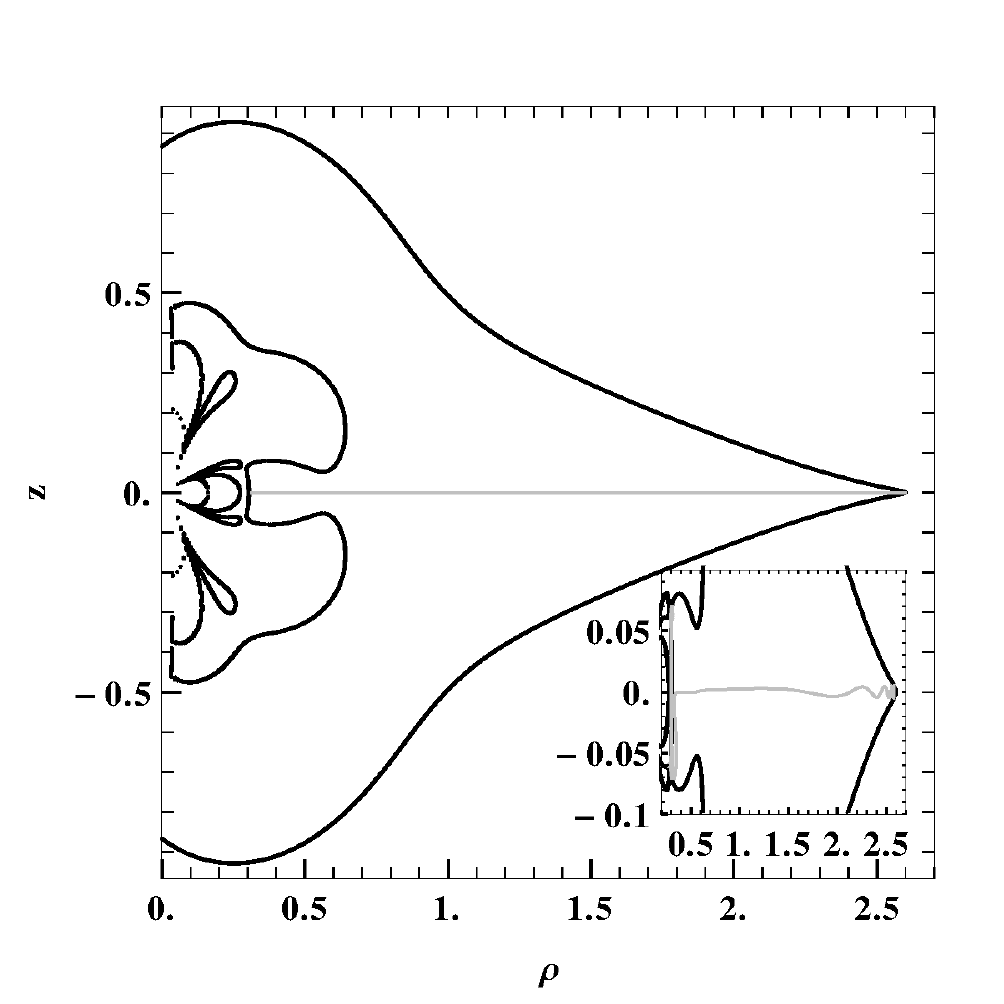}
 \caption{ The left panel shows the effective potential along $\rho$ on the equatorial plane
   ($z=0$) for the MN spacetime with parameters $M=1$, $a_*=0.5$, $q=-0.15$, and $h=-0.029$.
   The orbital parameters are $E=0.911388$ and $L_z=2.820706$, corresponding to 
   the ones of a circular orbit at the ISCO radius. For the same spacetime and orbital 
   parameters, in the right panel we show the CZVs (black curves) and an equatorial 
   orbit ($z=\dot{z}=0$) leaving the ISCO (the gray curve). In the embedded plot, 
   the gray curve shows a non-equatorial orbit leaving the ISCO.    
}
\label{FigISCO}
  \end{center}
\end{figure*}

Of course, our scenario can effectively work if the small body does jump from one
region to another, without plunging directly into the central object. The correct
approach to see if this is indeed the case is to compute the trajectory of one of
these bodies including the radiation reaction. However, there is not an established
prescription in the bibliography (at least to our knowledge) for handling radiation
reactions in non-Kerr spacetimes. There are some tricks one can use in
the relatively weak field limit~\cite{mn-glm,mn-glg}, but they cannot be used here, as
in our case we are considering the region close to the compact object, where gravity
is strong.
Moreover, when the small body leaves the inner edge of a region of stable
orbits, its trajectory is surely neither circular nor eccentric, and there might
even be a chance that the orbit gets non-equatorial due to vertical instabilities,
which makes the  issue of radiation reaction even more difficult to handle in the
strong regime. Therefore, in support of our scenario, we have simply checked if
(in the geodesic approximation) the small body plunges quickly into the central one
after having reached the ISCO radius or if it gets trapped in an inner region formed
around circular orbits. Our numerical calculations show that
the small body would indeed be trapped in an inner region, without plunging
directly to the central object.

In the specific example we examined, we used a MN spacetime with parameters $M=1$,
$a_*=0.5$, $q=-0.15$, $h=-0.029$. The effective potential on the
equatorial plane for energy and angular momentum corresponding to a
circular orbit at the ISCO radius (left panel
of Fig.~\ref{FigISCO}) shows that between the ISCO and the horizon ($\rho=0$) lies a
deep potential well, which indicates the existence of stable circular
orbits. On the $\rho-z$ plane in the right panel of Fig.~\ref{FigISCO}, the CZVs 
(black curves) show the allowed region for an orbit leaving the ISCO. 
The deep well of the $V_{\rm eff}$ corresponds to the loop lying in
the $0.3<\rho<0.6$ interval. If we evolve an equatorial geodesic orbit starting
near $\rho = \rho_{\rm ISCO}$, the orbit goes straight to the center of the well 
(gray curve in the right panel of Fig.~\ref{FigISCO}), which means that it will be 
trapped for a while in the inner region of circular orbits, before the final plunge. 
If we now evolve a non-equatorial geodesic orbit starting near the ISCO radius, the orbit
oscillates around the equatorial plane, but it also ends up in the well
(embedded plot in the right panel of Fig.~\ref{FigISCO}). Thus, even
some vertical instabilities may not cause a direct plunge and may allow
the small body to jump to an inner region of stable circular orbits.  
In the Kerr space-time, the particle cannot get trapped at small 
radii simply because there is no deep potential well between the object and the ISCO.

\begin{figure*}[htp]
  \begin{center}
  \includegraphics[type=pdf,ext=.pdf,read=.pdf,width=8cm]{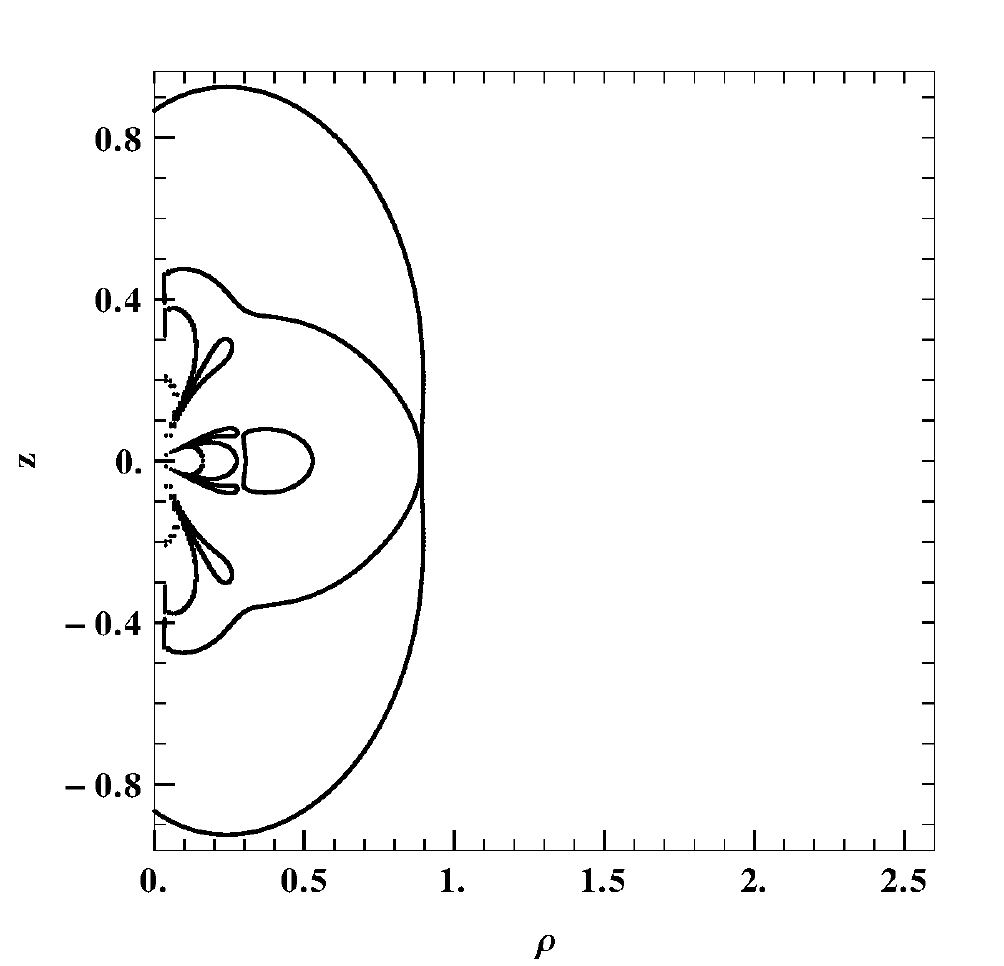}
  \includegraphics[type=pdf,ext=.pdf,read=.pdf,width=8cm]{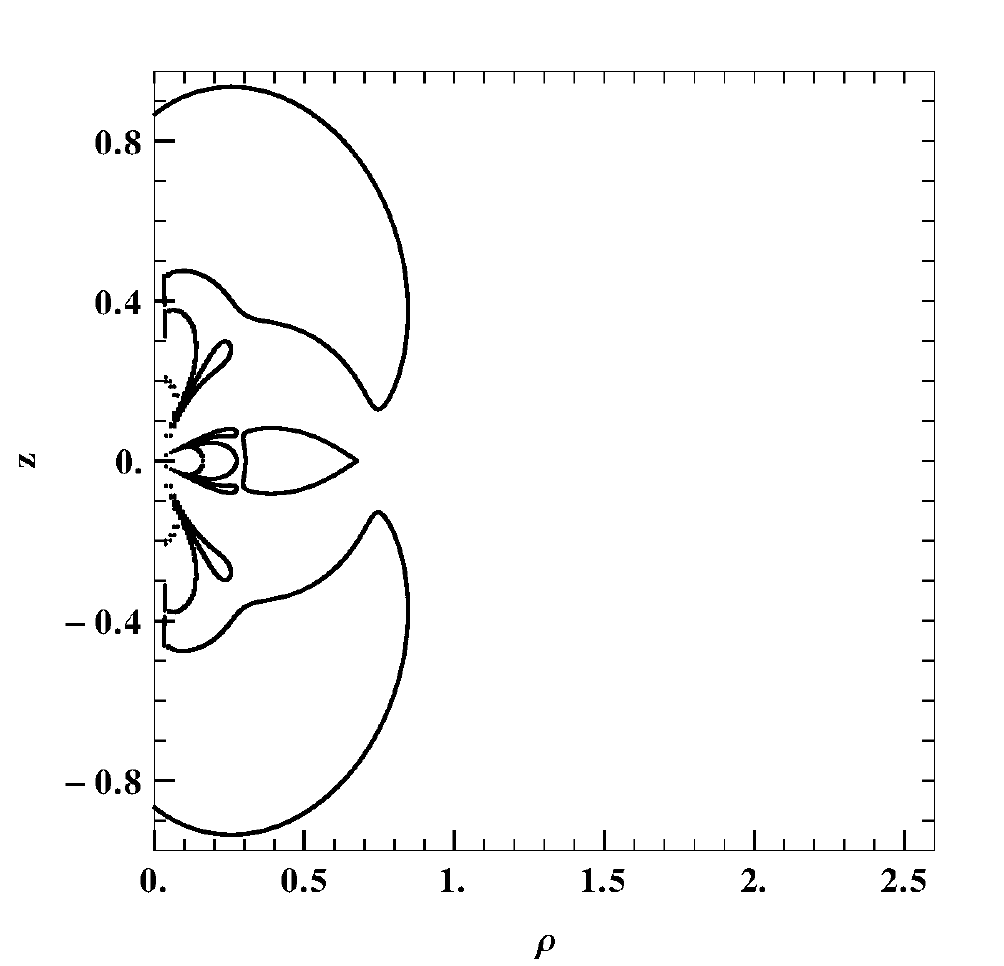}
 \caption{ The CZVs for the same spacetime parameters of Fig. \ref{FigISCO} but
 for different orbital parameters. The left panel corresponds to $E=0.888672$ and
 $L_z=2.773112$, and the right to $E=0.796835$ and $L_z=2.378980$.    
}
\label{FigCZVsco}
  \end{center}
\end{figure*}

If we reduce the levels of energy and angular momentum, as would happen due 
to the radiation reaction, we find two marginally stable circular orbits, 
see Fig.~\ref{FigCZVsco}. The first one is
radially stable and vertically unstable and has orbital parameters
$E=0.888672$ and $L_z=2.773112$. The second one is radially unstable and vertically
stable, with orbital parameters $E=0.796835$ and $L_z=2.378980$. 
As the stellar-mass body leaves the ISCO radius, it loses energy and angular momentum. 
This changes the CZVs. In particular, the large allowed region in Fig.~\ref{FigISCO} 
may split. Depending on how the small body loses energy and angular momentum,
it may adjust in the inner stable region at larger or smaller radii.

\section{Summary and conclusions}

Astrophysical BH candidates are thought to be the Kerr BHs predicted by General
Relativity, but this conclusion is not yet supported by robust observational
evidence. In this paper, we discussed the observational implications in the case 
that BH candidates are exotic compact objects whose exterior gravitational field is 
described by the Manko-Novikov spacetimes, which are exact solutions of the vacuum 
Einstein's equations. Such a possibility is not in contradiction with current 
observations, but it requires some form of exotic matter beyond standard physics. 
However, the basic scope of our work is to provide an observational test of
whether the centers of galaxies are occupied by Kerr BHs or by something 
more exotic, and not to argue that these exotic objects exist.

Equatorial circular orbits around a Kerr BH are always vertically stable, while
they are radially stable only for radii larger than $r_{\rm ISCO}$. The existence of
stable equatorial circular orbits at radii smaller than  $r_{\rm ISCO}$ in non-Kerr
spacetimes was already known, but so far they have never been considered to be of 
astrophysical interest as it was thought that their orbital energy was higher than 
the one at the ISCO and therefore that they could not be filled by the accretion 
gas or by small bodies inspiralling into the compact object. In this paper, we
have pointed out that this is not always true. We have also argued that the maximum 
number of these non-plunging regions may increase as the structure of the 
spacetime becomes more and more complex by adding higher anomalous mass 
moments. Lastly, we have proposed that the existence of these regions of stable 
equatorial circular orbits around astrophysical BH candidates may be tested by 
observing the gravitational waves emitted by a stellar-mass compact object
inspiralling into a super-massive BH candidate. In the standard case of a Kerr BH,
the frequency and the amplitude of the waveform should regularly increase, 
in the so-called ``chirp'' of the system, up to the plunging of the small body into 
the BH. In the presence of these inner regions, the total waveform could instead 
consist of ``regular chirps'', produced when the stellar-mass compact object
orbits in one of these stable regions, and ``bursts'', necessary to adjust the small 
body to a new stable orbit of a more internal region. Such a signature in the 
waveform might be observed with future space-based gravitational wave detectors.


\begin{acknowledgments}
We thank Enrico Barausse for reading a preliminary version of the 
manuscript and providing useful feedback.
C.B. was supported by the Thousand Young Talents Program and Fudan 
University. G.L.G. was supported by the DFG grant SFB/Transregio 7.
\end{acknowledgments}



\begin{thebibliography}{99}

\bibitem{hair} 
  B.~Carter,
  Phys.\ Rev.\ Lett.\  {\bf 26}, 331 (1971);
  D.~C.~Robinson,
  Phys.\ Rev.\ Lett.\  {\bf 34}, 905 (1975).

\bibitem{bh1} 
  R.~A.~Remillard and J.~E.~McClintock,
  Ann.\ Rev.\ Astron.\ Astrophys.\  {\bf 44}, 49 (2006)
  [astro-ph/0606352].

\bibitem{bh2} 
  J.~Kormendy and D.~Richstone,
  Ann.\ Rev.\ Astron.\ Astrophys.\  {\bf 33}, 581 (1995).
  
\bibitem{bh5} 
  M.~C.~Miller and E.~J.~MColbert,
  Int.\ J.\ Mod.\ Phys.\ D {\bf 13}, 1 (2004)
  [astro-ph/0308402].

\bibitem{bh3} 
  C.~E.~Rhoades, Jr. and R.~Ruffini,
  Phys.\ Rev.\ Lett.\  {\bf 32}, 324 (1974); 
  V.~Kalogera and G.~Baym,
  Astrophys.\ J.\  {\bf 470}, L61 (1996)
  [astro-ph/9608059].

\bibitem{bh4} 
  E.~Maoz,
  Astrophys.\ J.\  {\bf 494}, L181 (1998)
  [astro-ph/9710309].

\bibitem{review} 
  C.~Bambi,
  Mod.\ Phys.\ Lett.\ A {\bf 26}, 2453 (2011)
  [arXiv:1109.4256 [gr-qc]];
  C.~Bambi,
  Phys.\ Rev.\ D {\bf 85}, 043002 (2012)
  [arXiv:1201.1638 [gr-qc]];
  C.~Bambi,
  Phys.\ Rev.\ D {\bf 86}, 123013 (2012)
  [arXiv:1204.6395 [gr-qc]];
  C.~Bambi,
  Astron.\ Rev.\  {\bf 8}, 4 (2013)
  [arXiv:1301.0361 [gr-qc]].  
  
\bibitem{mn}
  V.~S.~Manko and I.~D.~Novikov,
  Class.\ Quant.\ Grav.\  {\bf 9}, 2477 (1992).
  
\bibitem{mn-glm} 
  J.~R.~Gair, C.~Li and I.~Mandel,
  Phys.\ Rev.\ D {\bf 77}, 024035 (2008)
  [arXiv:0708.0628 [gr-qc]].  
  
\bibitem{mn-glg} 
  T.~A.~Apostolatos, G.~Lukes-Gerakopoulos and G.~Contopoulos,
  Phys.\ Rev.\ Lett.\  {\bf 103}, 111101 (2009)
  [arXiv:0906.0093 [gr-qc]];
  G.~Lukes-Gerakopoulos, T.~A.~Apostolatos and G.~Contopoulos,
  Phys.\ Rev.\ D {\bf 81}, 124005 (2010)
  [arXiv:1003.3120 [gr-qc]].  
  
\bibitem{mn-cb} 
  C.~Bambi and E.~Barausse,
  Astrophys.\ J.\  {\bf 731}, 121 (2011)
  [arXiv:1012.2007 [gr-qc]];
  C.~Bambi,
  Europhys.\ Lett.\  {\bf 94}, 50002 (2011)
  [arXiv:1101.1364 [gr-qc]].
  C.~Bambi,
  Phys.\ Rev.\ D {\bf 83}, 103003 (2011)
  [arXiv:1102.0616 [gr-qc]];
  C.~Bambi,
  JCAP {\bf 1105}, 009 (2011)
  [arXiv:1103.5135 [gr-qc]];
  C.~Bambi,
  Phys.\ Rev.\ D {\bf 85}, 043001 (2012)
  [arXiv:1112.4663 [gr-qc]].

\bibitem{mn-cbeb} 
  C.~Bambi and E.~Barausse,
  Phys.\ Rev.\ D {\bf 84}, 084034 (2011)
  [arXiv:1108.4740 [gr-qc]];
  Z.~Li and C.~Bambi,
  JCAP {\bf 1303}, 031 (2013)
  [arXiv:1212.5848 [gr-qc]].
  
\bibitem{pqr} 
  D.~Pugliese, H.~Quevedo and R.~Ruffini,
ÊÊPhys.\ Rev.\ D {\bf 83}, 024021 (2011)
ÊÊ[arXiv:1012.5411 [astro-ph.HE]];
  D.~Pugliese, H.~Quevedo and R.~Ruffini,
ÊÊPhys.\ Rev.\ D {\bf 84}, 044030 (2011)
ÊÊ[arXiv:1105.2959 [gr-qc]].

 \bibitem{chl}
  G. Contopoulos, M. Harsoula and G. Lukes-Gerakopoulos,
  {\it Celest. Mech. Dyn. Astron.} {\bf 113}, 255 (2012);
  G. Lukes-Gerakopoulos and G. Contopoulos,
  (in preparation).
  
\bibitem{exotic1} 
  E.~K.~Boyda, S.~Ganguli, P.~Horava and U.~Varadarajan,
  Phys.\ Rev.\ D {\bf 67}, 106003 (2003)
  [hep-th/0212087];
  N.~Drukker,
  Phys.\ Rev.\ D {\bf 70}, 084031 (2004)
  [hep-th/0404239];
  E.~G.~Gimon and P.~Horava,
  hep-th/0405019.  

\bibitem{exotic2} 
  D.~Israel,
  JHEP {\bf 0401}, 042 (2004)
  [hep-th/0310158].
  
\bibitem{ergoregion} 
  P.~Pani, E.~Barausse, E.~Berti and V.~Cardoso,
  Phys.\ Rev.\ D {\bf 82}, 044009 (2010)
  [arXiv:1006.1863 [gr-qc]].
  
\bibitem{apj12} 
  C.~Bambi,
  Astrophys.\ J.\  {\bf 761}, 174 (2012)
  [arXiv:1210.5679 [gr-qc]]. 

\end{thebibliography}
\end{document}